\documentclass[preprintnumbers,prd,twocolumn,showpacs,letterpaper,nofootinbib]{revtex4-1}

\usepackage[pdftex]{graphicx}
\usepackage{latexsym,amsmath,amssymb,lmodern,float,url}
\usepackage{natbib}
\usepackage[pdftex,bookmarks,linktocpage,pdfpagelabels,plainpages=false,hyperfigures,linkcolor=blue,citecolor=blue]{hyperref} 
\hypersetup{colorlinks=true}

\newcommand\ee{\end{equation}}
\newcommand\be{\begin{equation}}
\newcommand\eea{\end{eqnarray}}
\newcommand\bea{\begin{eqnarray}}

\begin{document}
\preprint{MI-TH-1618}
\preprint{CETUP2016-003}

\title{Dark matter, light mediators, and the neutrino floor}
\author{James B.~Dent$^{\bf ab}$}
\author{Bhaskar Dutta$^{\bf c}$}
\author{Jayden L.~Newstead$^{\bf d}$}
\author{Louis E.~Strigari$^{\bf c}$}

\affiliation{$^{\bf a}$ Department of Physics, University of Louisiana at Lafayette, Lafayette, LA 70504, USA}
\affiliation{$^{\bf b}$ Kavli Institute for Theoretical Physics, University of California, Santa Barbara, CA 93106-4030, USA}
\affiliation{$^{\bf c}$ Mitchell Institute for Fundamental Physics and Astronomy,  Department of Physics and Astronomy, Texas A\&M University, College Station, TX 77845, USA}
\affiliation{$^{\bf d}$ Department of Physics, Arizona State University, Tempe, AZ 85287, USA}
 
\begin{abstract}
We analyze future direct data matter detection experiments using Effective Field Theory (EFT) operators with light,  $\lesssim 100$ MeV mass mediators. We compare the nuclear recoil energy spectra from these operators to the predicted high energy solar neutrino spectrum. A set of operators that generate spectra similar to the neutrino background is identified, however this set is distinct from those that mimic the neutrino background for heavy,  $\gtrsim 100$ MeV mass mediators. We outline a general classification scheme based on momentum dependence of the dark matter-nucleus interaction to determine how strong the discovery limit for a given operator saturates in the presence of the neutrino background.   Our results highlight the benefit of considering a general theoretical framework regarding dark matter, and motivate continued experimental progress towards lower nuclear recoil energy thresholds. 
\end{abstract}

\maketitle

\par \emph{Introduction.}\textemdash Direct dark matter detection experiments continue to extend their sensitivity reach to lower scattering cross-sections~\cite{XENON100,Akerib:2015rjg}, as well as extension towards lower energy thresholds~\cite{Agnese:2015nto}. The development of multi-ton scale detectors which will deeply explore the weakly interacting massive particle (WIMP) mass and scattering cross-section parameter space~\cite{Aprile:2015uzo,Akerib:2015cja,Aalbers:2016jon} calls for a clear understanding of how to properly connect dark matter models with any future observations~\cite{Newstead:2013pea,Peter:2013aha}.

\par In recent years theoretical modeling of WIMP-nucleus ($\chi$-$N$) scattering has expanded beyond the point-nucleus framework~\cite{Lewin:1995rx}, in which the nucleus is treated as a point particle and momentum dependence is encoded via a form factor, giving rise to spin-independent (SI) and spin-dependent (SD) interactions where the SI interaction is enhanced due to a coherence effect at low momentum transfer.  An enlarged view of modeling $\chi$-$N$ scattering beyond this standard SI/SD framework has emerged~\cite{Fan:2010gt,Fitzpatrick:2012ix,Anand:2013yka}. This current state-of-the art is the implementation of a general non-relativistic effective field theory (EFT) approach which encompass a wide variety of interactions including general velocity and momentum dependence, as well as inclusion of additional nuclear responses beyond those employed in the traditional SI and SD cases.

\par As the sensitivity of direct searches increases, an important background is expected to arise from the coherent neutrino-nucleus scattering from astrophysical neutrino sources.  In the usual SI/SD scenario, distinguishing $\nu$-$N$ scattering from $\chi$-$N$ scattering will prove challenging.  For example, in the limit where the mass of the mediating particle is large compared to the momentum transfer of the scattering process, $\chi$-N scattering for WIMP masses $\sim$ 100 GeV, 6 GeV, and 1 GeV, are degenerate with atmospheric, $^8$B solar neutrinos, and $^7$Be solar neutrinos, respectively~\cite{Billard:2013qya}. Therefore, once direct detection experiments become sensitive to coherent neutrino scattering from these sources, the prospects for dark matter detection in the near future could diminish considerably.

\par However, prospects for identifying a dark matter signal in light of the neutrino backgrounds depends on the nature of the physics that governs dark matter-nucleus interactions. Ref.~\cite{Dent:2016iht} recently showed that the discovery potential for the majority of the general EFT interactions (encoded in the form of fifteen $\chi$-$N$ operators and six nuclear responses) shows little abatement in the presence of neutrino backgrounds due to the form of their momentum and velocity dependences, implying that the neutrino background can be distinguished in future experiments.  The $\chi$-$N$ interactions in Ref.~\cite{Dent:2016iht} were assumed to be mediated by scalar and vector particles whose masses are well above the standard momentum transfer scale of $|\vec{q}| \sim 100$MeV, with the four-momentum $q$ equivalent to $-\vec{q}$ in the non-relativistic limit which is applicable to $\chi$-$N$ scattering.  

\par In this paper we relax this assumption and include mediators (here generically referred to as $\phi$ for both the scalar and vector case) with masses $m_{\phi}^2 \lesssim |\vec{q}|^2$. Models for light particles which couple the Standard Model to new hidden sectors through such light mediators have been developed, for a recent review see Ref.~\cite{Essig:2013lka}, and for recent work on direct detection with light mediators see Ref.~\cite{Li:2014vza,DelNobile:2015uua}. In addition Ref.~\cite{Cerdeno:2016sfi} explores non-standard neutrino interactions which may affect coherent scattering. For our purposes, we are interested in the fact that a light mediator will alter the momentum dependence of the differential scattering cross-section as the denominator of the propagator will no longer have the limiting $(m_{\phi}^2 + \vec{q}^2) \rightarrow m_{\phi}^2$ as is the case for a heavy mediator. As we show, the set of operators which are distinguishable from the neutrino background is different than the set for the heavy mediator case. Therefore the light mediator scenario implies unique phenomenology in upcoming direct dark matter searches, including the possibility of extracting an SI/SD signal in the presence of the neutrino background.

\par\emph{EFT formalism}\textemdash The complete set of non-relativistic operators arising from the reduction of a relativistic treatment, which describes elastic $\chi$-$N$ scattering due to spin-0 or spin-1 mediator exchange up to second order in momentum, is comprised of ten operators~\cite{Fan:2010gt,Fitzpatrick:2012ix,Anand:2013yka}. There exist four additional operators that can also be written down at this order which do not arise from traditional` single mediator exchange~\cite{Anand:2013yka}. All fourteen of these operators are written in terms of four quantities: the exchanged momentum, $\vec{q}$, the $\chi$-$N$ relative incident velocities $\vec{v}$ in the form of the variable $\vec{v}^{\perp} = \vec{v} + \vec{q}/2\mu_N$ with $\mu_N$ the $\chi$-$N$ reduced mass, the spin of the dark matter $\vec{S}_{\chi}$, and the nucleon spin $\vec{S}_N$ (there are actually fifteen operators that arise at this order, but one operator is proportional to $(\vec{v}^{\perp})^2$ which does not appear as the NR reduction of a relativistic operator, and is not considered here).  

\par These fourteen operators can further be categorized into three groups which display similar momentum and dark matter lab frame velocity ($v_T$) dependence \cite{Catena:2015vpa,Gluscevic:2015sqa,Dent:2016iht}. Group I operators have no $q^2$ dependence, Group II have $q^2$ and  $q^2v_T^2$ dependence, while Group III have $q^2v_T^2$, $q^4$, and $q^4v_T^2$ dependences. This momentum and velocity dependence is obtained in the limit where the masses of the mediator particles are large compared to the momentum transfer of the interaction. In the presence of the neutrino backgrounds, operators from each group display similar discovery evolution limits, and a similar dark matter mass mimcs each group~\cite{Dent:2016iht}.  

\par When mediator masses $\lesssim |\vec{q}|$ are considered, the same group structure can be used, since their relative momentum dependences are the same. However, the important distinction is that the {\em overall momentum dependence within each group is different than in the case of heavy mediators.} As we see below this has drastic consequences for which group of operators can be distinguished from the neutrino background.  In the non-relativistic Lagrangian each operator has a dimensionful coupling, $c_i$, previously taken to be proportional to $1/m_v^2$. To encapsulate the light mediator scenario we employ the replacement,
\be
c_i \rightarrow \frac{c_i}{q^2+m_\phi^2},
\ee
where $c_i$ is now a dimensionless constant. Given the low momentum transfer from WIMPs to the nuclei, a mediator mass $\gtrsim100$ MeV will dominate over the $q^2$ term in the propagator. Therefore we will consider three scenarios: mediators of mass 1 MeV, 10 MeV and 100 MeV, which correspond to the scenario with $q^2 > m_\phi^2$, $q^2\sim m_\phi^2$ and $q^2 < m_\phi^2$. 

\par\emph{Discovery Evolution}\textemdash The distinguishability of operators in the presence of the coherent neutrino scattering background amounts to examining whether the discovery evolution as a function of the detection exposure (the product of target mass and time) saturates. This corresponds to a situation in which increasing detector exposure is ineffective at extending the discovery reach to lower $\chi$-$N$ cross-sections~\cite{Billard:2013qya}. Eventually enough statistics could be compiled that would end the saturation effect, but once the saturation occurs it persists for several orders of magnitude of exposure, thus nullifying any practical chances of discovery once saturation has been reached~\cite{Ruppin:2014bra}. In the EFT framework with heavy mediators, two of the three groups, equating to ten out of the fourteen operators, do not experience such a saturation effect, and therefore could possibly be distinguished even in the presence of a background of coherent neutrino scattering~\cite{Dent:2016iht}.  

\par Here we examine the representative operators $\mathcal{O}_1$, $\mathcal{O}_{10}$, and $\mathcal{O}_{6}$ from Group I ($\mathcal{O}_{1,4,7,8}$), II ($\mathcal{O}_{5,9,10,11,12,14}$), and III ($\mathcal{O}_{3,6,13,15}$), respectively, in the case of mediators $m_\phi < 100$ MeV.  $\mathcal{O}_1$ is the standard SI operator which, along with the rest of Group I operators, exhibits a saturated discovery evolution in the case of heavy mediators. By comparison Group II and III operators do not exhibit this saturation for heavy mediators. The $\mathcal{O}_6$ operator has the NR form $(\vec{S}_{\chi}\cdot\vec{q}/m_N)(\vec{S}_N\cdot\vec{q}/m_N)$ and arises in dipole interacting dark matter and pseudoscalar mediated interactions, and $\mathcal{O}_{10} = i\vec{S}_N\cdot\vec{q}/m_N$ also arises in pseudoscalar mediated scattering. Note that these operators can be connected to various scattering models~\cite{Gresham:2014vja,Dent:2015zpa,Gluscevic:2015sqa}.

\par We begin by matching the nuclear recoil spectra from the various WIMP-nucleon operators described above to the predicted $^8$B solar neutrino-induced recoil energy spectrum, for various mediator masses. To obtain the predicted recoil energy spectra in dark matter detectors due to these neutrinos, we use the high metallicity standard solar model predictions, e.g.~\cite{Strigari:2016ztv}. For a heavy mediator, the $^8$B rate is well-fit by SI interacting dark matter with a mass of $m_{\chi} \simeq 6$ GeV. To find this ``best-fit" WIMP masses for any given operator we maximize the Poisson likelihood,
\be
\mathcal{L}_{Poisson} = \prod_{i=1}^b \frac{\nu_i^{n_i} e^{-\nu_i}}{n_i!}
\label{eq:Lpoisson} 
\ee
where $b$ is the number of nuclear recoil energy bins, $n_i$ is the expected number of WIMP events and $\nu_i$ is the expected number of neutrino events in the bin. To demonstrate the effect of light mediators on the discovery evolution we consider a single germanium detector with a threshold of 100 eV. We consider germanium as an example because it is an appropriate target to highlight the potential for $\sim 100$ eV low threshold recoil detectors. Our numerical results would be very similar if we were to instead consider a xenon target. For our likelihood analysis we choose an exposure such that we obtain 200 neutrino events for each target~\cite{Billard:2013qya}, binned into 16 energy bins.

\par The resulting best fit masses are given in Table~\ref{tabMasses}, where the masses are averaged between fits to neutron and proton rates (which do not differ significantly). For most groups, we find a reasonable correspondence between the neutrino and best fitting WIMP spectra. The main outlier is the case of $\mathcal{O}_1$ with a very light mediator. When performing the fit with $\mathcal{O}_1$ and a 1 MeV mediator the best fit is found at large WIMP mass, however the likelihood function plateaus in this limit. While all fits above $10^6$ GeV maximize the likelihood, the quality of the fit remains poor.

\begin{table}
\caption{Best fit WIMP masses, in GeV, to the $^8$B neutrino rates in germanium for various operators and mediator masses.}
\begin{tabular}{|l|c|r|r|r|}
\hline
 Operator                     & $q$ dependence  & \multicolumn{3}{|c|}{mediator mass ($m_{\phi}$)}\\
 \cline{3-5}
& & 100MeV & 10MeV & 1MeV\\
\hline
$\mathcal{O}_1$ (Group I)    &  $1$            & 6.3    &  13  &  $>10^6$  \\
$\mathcal{O}_{10}$ (Group II)&  $q$            & 5.6    &  6.5 &  12  \\  
$\mathcal{O}_6$ (Group III)  &  $q^2$          & 5.0    &  5.3 &  6.3   \\  
\hline
\end{tabular}
\label{tabMasses}
\end{table}


\par
\begin{figure*}[ht]
\begin{tabular}{ccc}
\includegraphics[height=4.5cm]{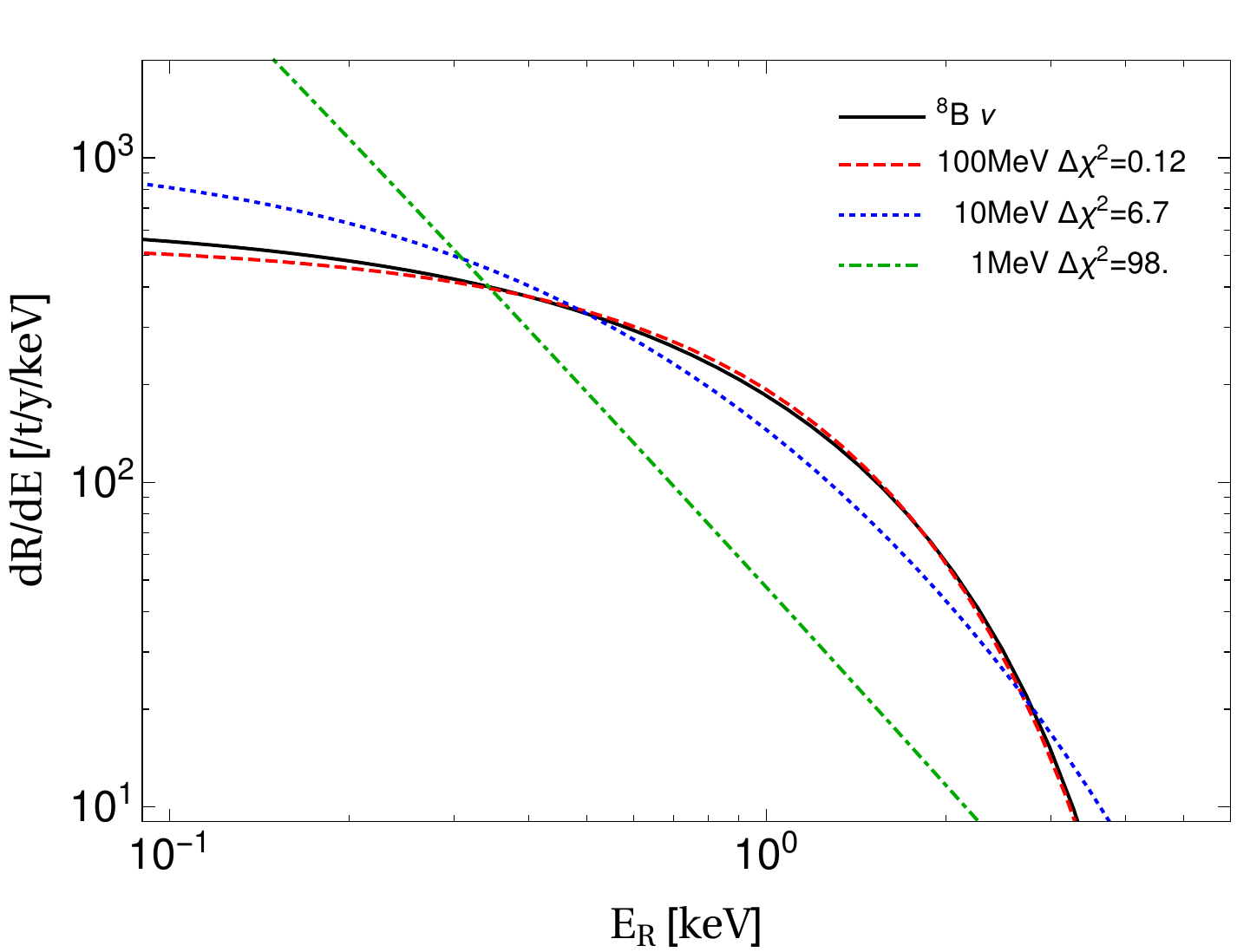}  &
\includegraphics[height=4.5cm]{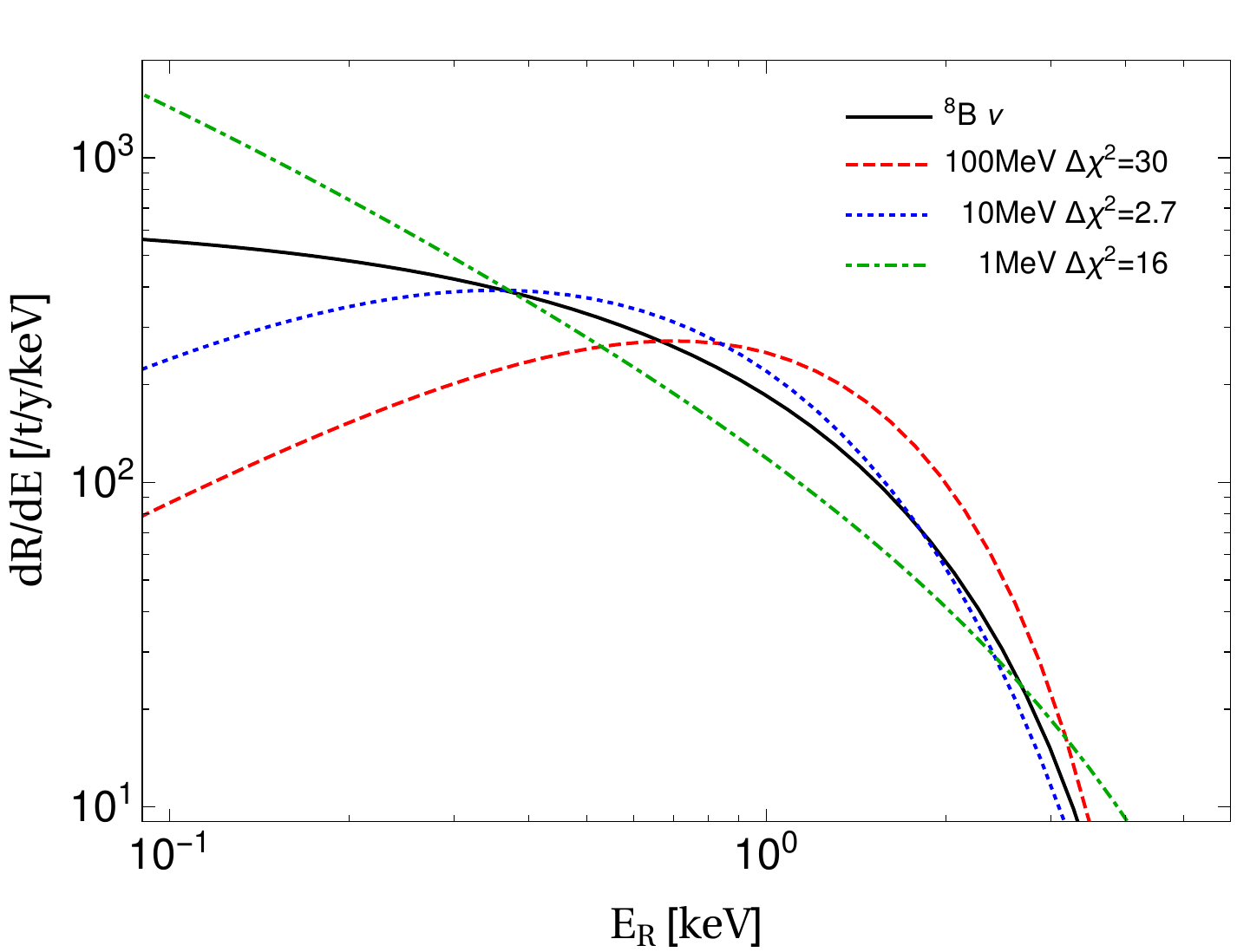}  &
\includegraphics[height=4.5cm]{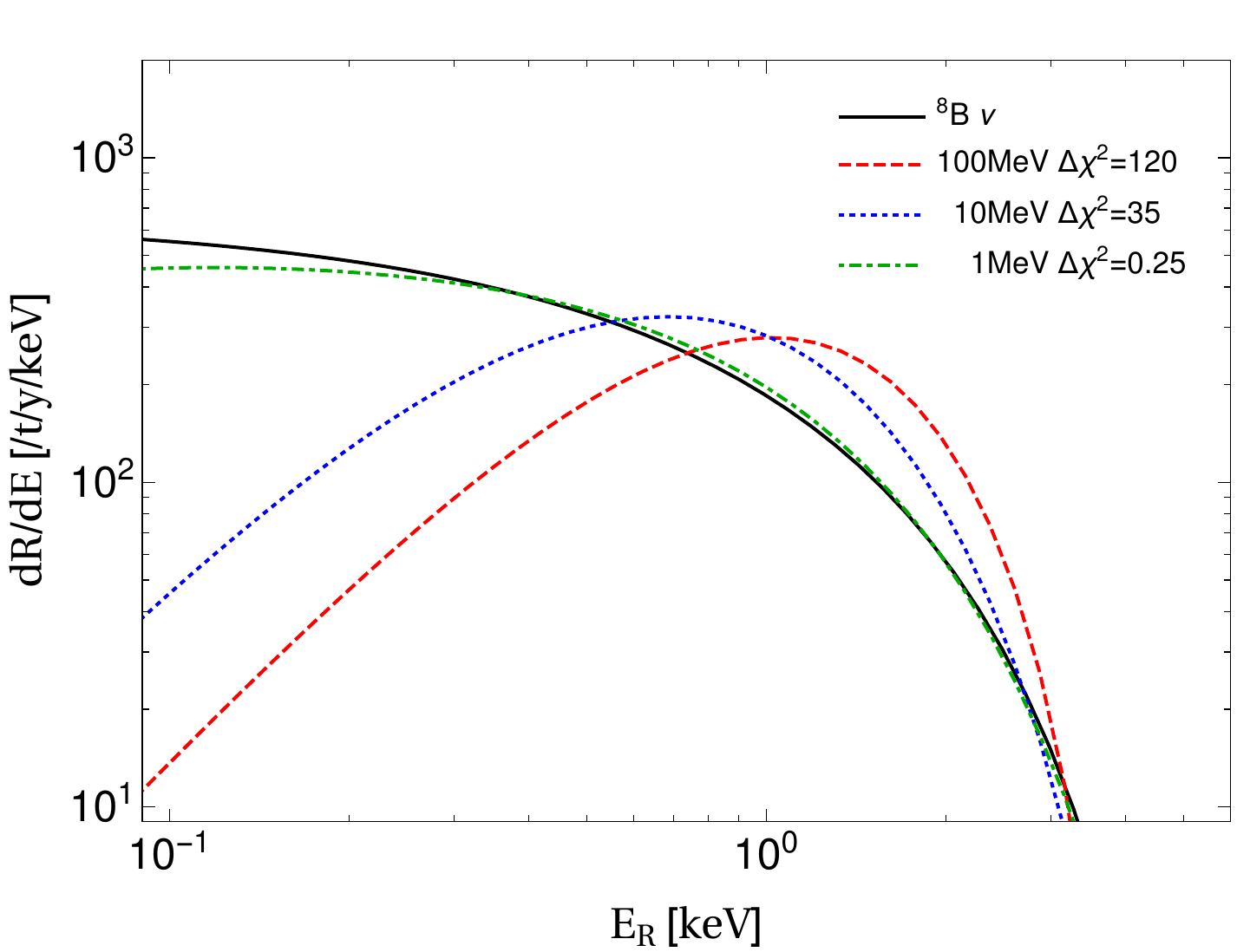} \\
\end{tabular}
\caption{Best fit recoil spectra fitted to $^8$B neutrino rates in germanium for $\mathcal{O}_1$ (left), $\mathcal{O}_{10}$ (middle) and $\mathcal{O}_6$ (right).  The solid black line displays the spectrum for coherent neutrino scattering, while the other curves denote different mediator masses.}
\label{fig:dRdE}
\end{figure*}

 The recoil spectra for the best fit masses are displayed in Figure~\ref{fig:dRdE}. This figure shows that for the case of light mediators, Groups I and II for $m_{\chi} =$ 6 GeV are poor fits to the $^8$B neutrino spectra, whereas Group III operators can fit it well, with the exception of $\mathcal{O}_{15}$. It should be emphasized that the deviation between the WIMP and neutrino spectra shows up most starkly at very low recoil energy, which provides good motivation for the development of low threshold detector technology~\cite{Mirabolfathi:2015pha}. Since $q^2$ is proportional to $v^2$, the full propagator in the numerator of the operator modifies the Group I rate to no longer be velocity independent, making it a poor fit to the neutrino background. The opposite is true for Group II and III, which can provide better fits to the neutrino background at low mediator mass.

\par To calculate the discovery potential, we follow the statistical formalism of Ref.~\cite{Billard:2013qya}. Recall that the discovery potential of an experiment is defined as the smallest WIMP-nucleon cross section which produces a 3$\sigma$ fluctuation above the background 90\% of the time. To calculate this limit we use the following test statistic for the null hypothesis and try to reject it,
\be
q_0 =
\begin{cases}
   -2 \mathrm{log}  \frac{\mathcal{L}(\sigma=0,\hat\theta)}{\mathcal{L}(\hat\sigma,\hat{\hat\theta})}  & \sigma  \geq \hat\sigma \\
   0	& \sigma < \hat\sigma \\
\end{cases}
\ee
where $\sigma$ is the WIMP-nucleon cross section, $\theta$ represents the nuisance parameters (neutrino fluxes), and the hatted parameters are maximized. By Wilks' theorem, under background only experiments, $q_0$ is chi-square distributed and the equivalent gaussian significance is simply $\sqrt{q_0}$~\cite{Cowan:2010js}.  To include the uncertainty of the neutrino flux normalization the likelihood function is modified to include a gaussian term~\cite{Billard:2013qya}:
\be
\mathcal{L} = \mathcal{L}_{Poisson} e^{-\frac{1}{2}(1-N_\nu)^2\left(\frac{\phi_\nu}{\sigma_\nu}\right)^2}
\ee
where $N_\nu$ is the flux normalization and $\phi_\nu=5.58\times10^6$ cm$^{-2}$ s$^{-1}$ and $\sigma_\nu=0.14\times10^6$ cm$^{-2}$ s$^{-1}$ are the $^8$B flux and uncertainty respectively. The poisson likelihood $\mathcal{L}_{Poisson}$ is defined as in Equation~\ref{eq:Lpoisson}. 

\par The ``worst case" scenario of the discovery evolution is where the WIMP spectrum most closely resembles the neutrino background. For combinations of operators and mediator masses which are sufficiently neutrino like, the evolution of the discovery potential exhibits saturation when the systematic uncertainty in the neutrino flux becomes relevant. This saturation is then broken when the exposure becomes large enough that small differences in the WIMP and neutrino-induced recoil spectra become distinguishable~\cite{Ruppin:2014bra}. For combinations of operators and mediator masses with recoil spectra that are sufficiently different than the neutrino-induced recoil spectra, no significant saturation is observed. For these cases a weak inflection point defines the exposure at which the saturation is a maximum. The scenarios that reach an inflection point at lower exposures are those that are most easily distinguishable from the neutrino backgrounds. These scenarios return to a $1/\sqrt{MT}$ evolution as the exposure is increased. 

\par
\begin{figure*}[ht]
\begin{tabular}{ccc}
\includegraphics[height=4.5cm]{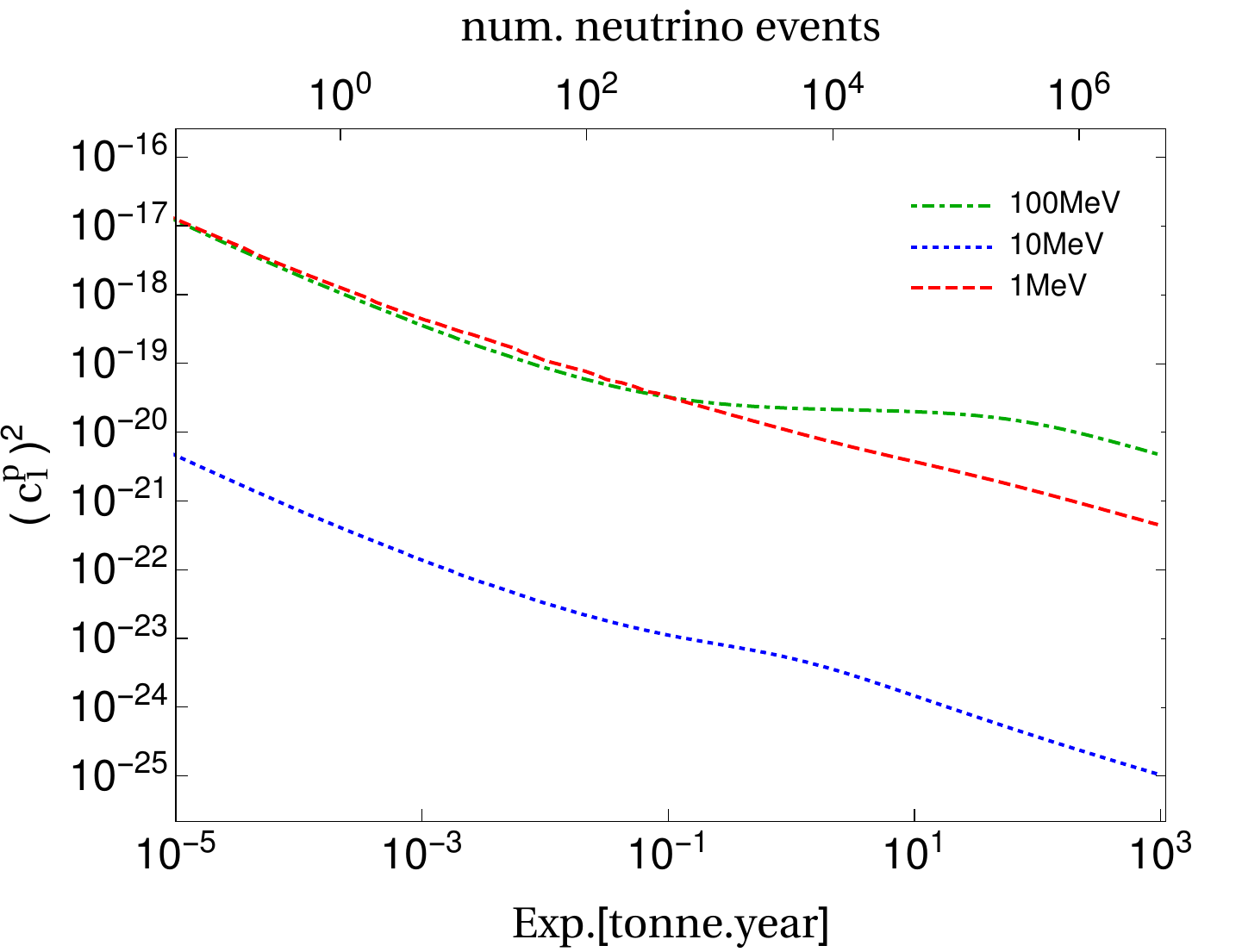}  &
\includegraphics[height=4.5cm]{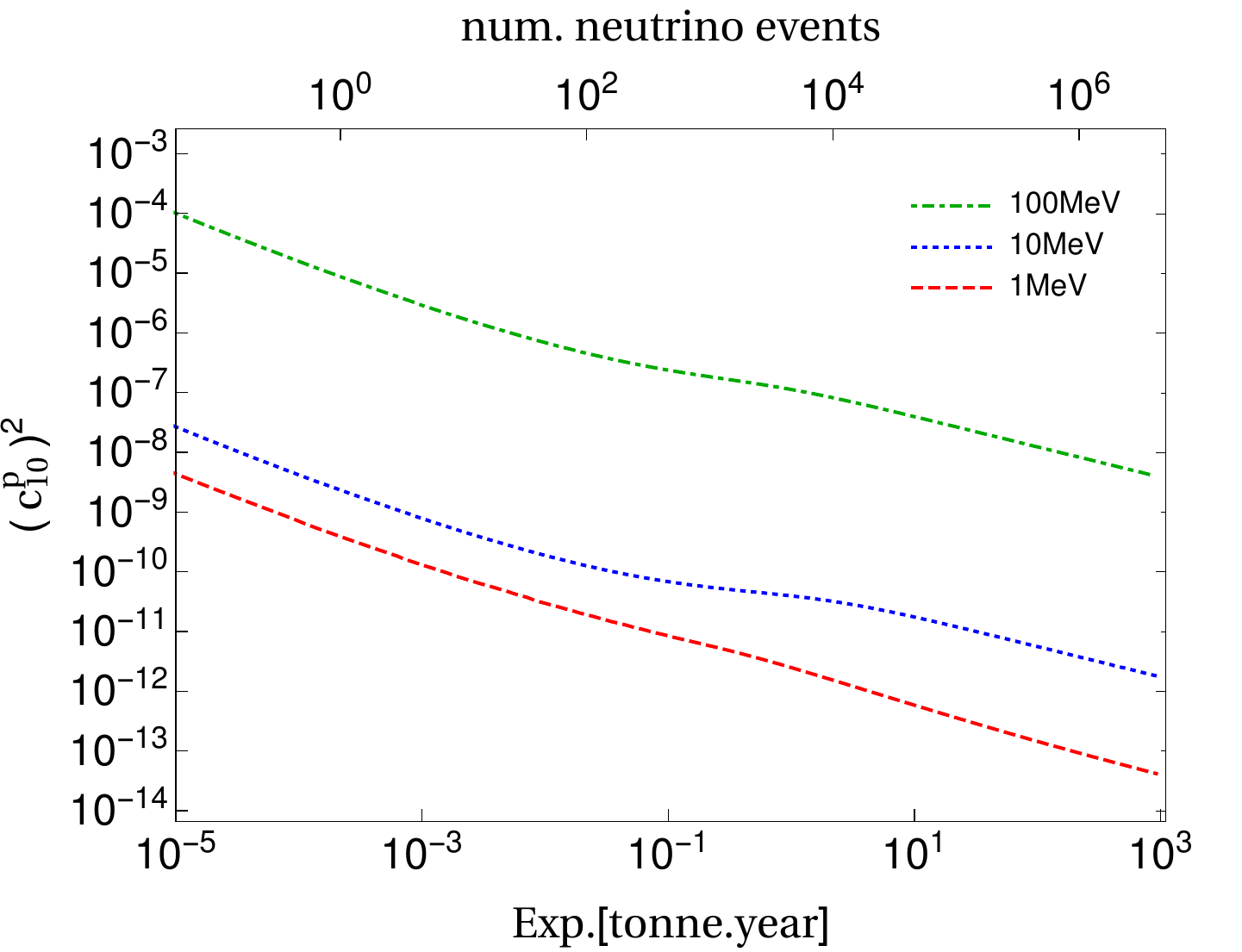}  &
\includegraphics[height=4.5cm]{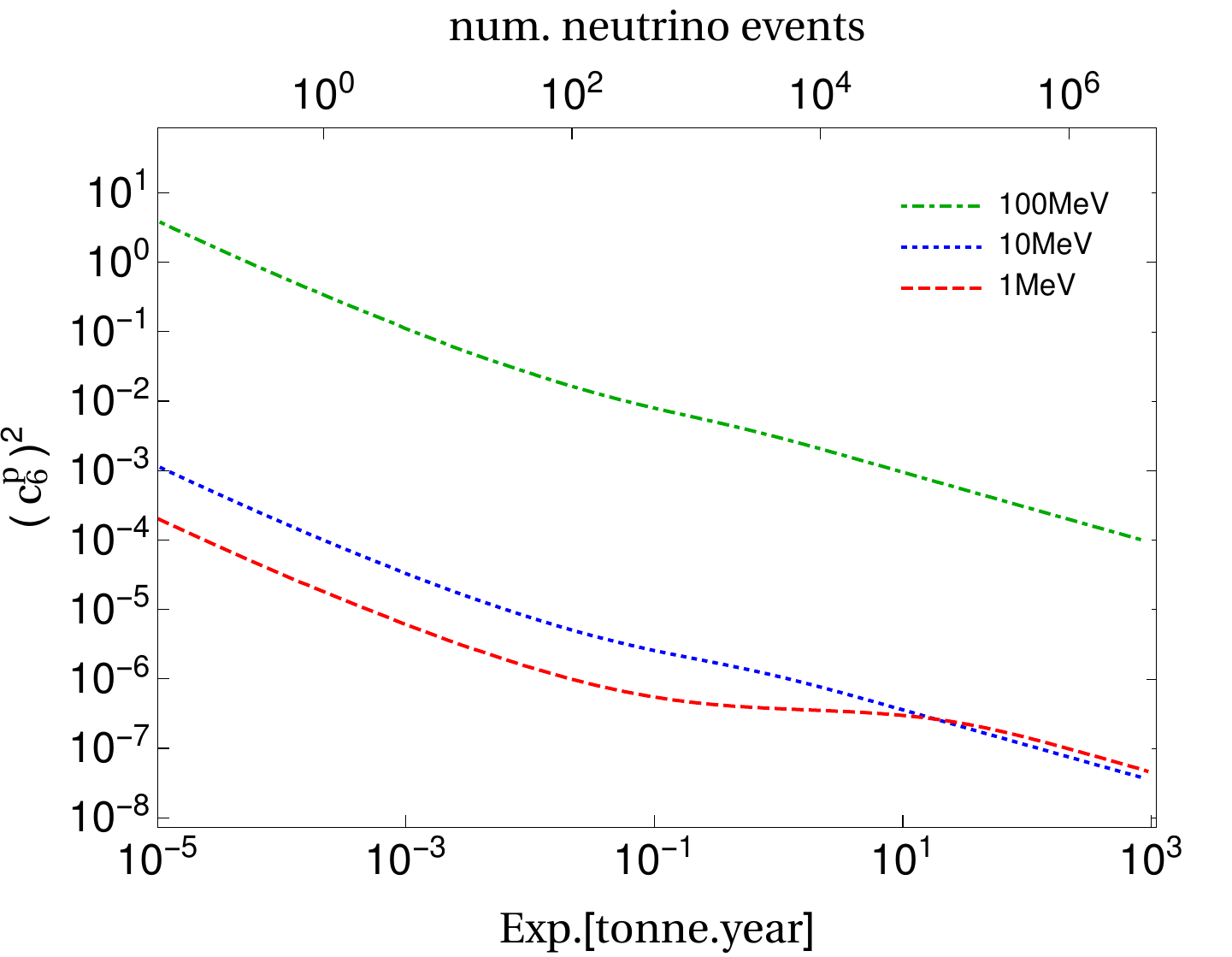} \\
\end{tabular}
\caption{Discovery evolution of $\mathcal{O}_1$ (left), $\mathcal{O}_{10}$ (middle) and $\mathcal{O}_6$ (right). The curves show the limits for proton scattering only}
\label{figDiscEvo}
\end{figure*}

 We calculate the evolution of the discovery potential for $\mathcal{O}_1$, $\mathcal{O}_6$ and $\mathcal{O}_{10}$ operators using a germanium based experiment, for the best fit WIMP mass to the $^8$B neutrino background (see Table~\ref{tabMasses}). This discovery evolution for $\mathcal{O}_1$, $\mathcal{O}_6$ and $\mathcal{O}_{10}$ for scattering off protons is shown in Figure~\ref{figDiscEvo}. The observed variation in the magnitude of the discovery reach for the different mediator masses is due to the best fit masses being different for each mediator mass. The corresponding neutron scattering evolution (not shown) is scaled by a constant factor. The discovery evolution for $\mathcal{O}_1$ saturates in the high mediator mass regime, less strongly with $m_\phi = 10$ MeV mediator, and hardly at all for $m_\phi = 1$ MeV. The reverse is observed for $\mathcal{O}_6$ which does not saturate with high mediator mass, however at low mediator mass it can mimic the neutrino rate. The $\mathcal{O}_1$ operator for mediator masses 10 and 1 MeV can be distinguished from the neutrino background by 0.1 ton years exposure using a Ge detector, whereas the 100 MeV or larger mediator mass requires $10^2$ ton years exposure.  The $\mathcal{O}_{10}$ operator can be distinguished by 10 ton years exposure or less for any mediator mass. For $\mathcal{O}_{6}$, mediator masses of 1 MeV and below require 10 ton years of exposure.

\begin{table}
\caption{Summary of whether saturation in the discovery evolution is observed for the various WIMP scattering scenarios}
\begin{tabular}{|c|c|c|}
\hline
 Group  & light mediator & heavy mediator \\
 & $m_{\phi} \lesssim 100$MeV & $m_{\phi} \gtrsim 100$MeV\\
\hline
Group I       & No       &  Yes \\
Group II      & No       &  No  \\ 
Group III     & Yes      &  No  \\
\hline
\end{tabular}
\label{tabFloors}
\end{table}

\par\emph{Conclusions}\textemdash We have shown that the character of the discovery potential for elastic dark matter scattering off of nuclei in the presence of the neutrino background greatly depends not only on the type of interaction, but also on the mass of the particle mediating the scattering process.  Table \ref{tabFloors} details for which operators, mediator masses and low mass dark matter particles the saturation of the discovery evolution for $\chi$-$N$ scattering persists, i.e. hits a neutrino floor. Interestingly even the standard SI and SD operators may be distinguishable for light mediators at a very low threshold detector, which was not the case for heavy mediators.  Conversely, some operators which were thought to be distinguishable from the neutrino background can be rendered indistinguishable for the same exposure when the mediator mass is sufficiently light.

These results demonstrate the necessity of considering a general theoretical framework regarding dark matter scattering when projecting future discovery potential, as well as increased motivation for experimental progress towards lower thresholds.

\par\emph{Acknowledgements} \textemdash 
BD and LES acknowledge supports from DOE Grant DE-FG02-13ER42020 and NSF grant PHY-1522717 respectively.

\bibliography{PhysicsBibtex} 

\end{document}